\documentclass[twocolumn,twoside,slac_two]{revtex4}
\usepackage{graphicx}
\usepackage{fancyhdr}
\usepackage{multirow}
\newcommand{\ee}{e^{+} e^{-}}

\newcommand{\jp}{J/\psi}
\newcommand{\psip}{\psi '}
\newcommand{\jpsi}{J/\psi}

\newcommand{\pipi}{\pi^{+}\pi^{-}}

\newcommand{\rt}{\rightarrow}

\newcommand{\dnd}     {\ensuremath{ D^0    \overline D {}^{0}}}
\newcommand{\dndst}   {\ensuremath{ D^0    \overline D {}^{*0}}}

\begin{document}

\title{Hadron Spectroscopy Results from Belle}

%

\author{A. Zupanc (for the Belle Collaboration)}
\affiliation{Karlsruhe Institute of Technology, 76131 Karlsruhe, Germany}

\begin{abstract}
We report on the most recent experimental progress on $XYZ$ charmoniumlike meson states and their possible counterparts in the $s\overline{s}$ and $b\overline{b}$ systems.
\end{abstract}

\maketitle

\thispagestyle{fancy}


\section{Introduction}
The results presented here are based on a data sample collected with the Belle detector~\cite{unknown:2000cg} at the KEKB asymmetric-energy $e^+e^-$ collider~\cite{Kurokawa:2001nw}. The experiment is located at the High Energy Accelerator Research Organization (KEK) in Tsukuba, Japan. Electrons and positrons are being collided at the center-of-mass (CM) energy of about 10.58~GeV, corresponding to the mass of the $\Upsilon(4S)$ resonance. The main objective of the Belle experiment is a study of the $CP$ violation in the $B$ meson system produced from $\Upsilon(4S)$, however a large data sample collected with the Belle detector and its excellent performance also make possible to perfrom searches for new hadronic states as well as studies of their properties. There are several possible mechanisms of the particle production at $B$ factories: production in the $B$ meson decays, fragmentation of quarks in $e^+e^-$ annihilation or two-photon collisions. Most of the newly discovered states, the so-called $XYZ$ mesons, almost certainly contain a $c\overline{c}$ quark pair among their constituent particles but do not fit into any of the unassigned charmonium levels predicted within the potential models. As a result, they are considered to be possible candidates for new, exotic types of particles such as multiquark states that include either molecular states (two loosely bound charm mesons ($c\overline{q}$)($\overline{c}q$)) \cite{molecular,molecular1,molecular2,molecular3,molecular4,molecular5,molecular6,molecular7,molecular8,molecular9},
or tetraquarks (tightly bound four-quark states, [$c\overline{q}$][$\overline{c}q$]) \cite{diquark,diquark1,diquark2,diquark3}, charmonium hybrids
($c\overline{c}$-gluon with excited gluonic degrees of freedom)~\cite{hybrid,hybrid1,hybrid2}, or
hadro-charmonium (compact charmonium states, $J/\psi$, $\psi(2S)$,
$\chi_c$, "coated" by excited light-hadron matter)~\cite{hadroch}. More conservative models \cite{thresh,thresh1,thresh2,thresh3} suggest to reconsider the effect of the numerous open charm thresholds which are located in many cases near the new states on the parameters of the conventional charmonium levels.

In this paper we review review recent results from Belle collaboration on the $X$, $Y$ and $Z$ charmoniumlike meson states as well as on their possible counterparts in $s\overline{s}$ and $b\overline{b}$ systems, summarized also in Table~\ref{tab_newstates}. 
\begin{table*}[t]
\caption{Summary of new states observed by Belle \cite{OlsenProc}.}
\label{tab_newstates}
\begin{tabular}{lllllll}
\hline\hline
& & & & & & Also\\
\multirow{-2}*{State}    & \multirow{-2}{*}{$M$~(MeV)} & \multirow{-2}{*}{$\Gamma$~(MeV)}    & \multirow{-2}{*}{$J^{PC}$} & \multirow{-2}{*}{Decay Modes}        & \multirow{-2}{*}{Production  Modes} & \multirow{-1}{*}{observed by } \\\hline

& & & & &  $\ee$~(ISR) & \\
\multirow{-2}{*}{$Y_s(2175)$}& \multirow{-2}{*}{$2175\pm8$} & \multirow{-2}{*}{$ 58\pm26 $} & 
\multirow{-2}{*}{$1^{--}$} & \multirow{-2}{*}{$\phi f_0(980)$}     & $\jp\rt\eta Y_s(2175)$ &  \multirow{-2}{*}{BaBar, BESII}\\

&&&&$\pipi\jp$, & & BaBar\\
\multirow{-2}{*}{$X(3872)$}& \multirow{-2}{*}{$3871.4\pm0.6$} & \multirow{-2}{*}{$<2.3$} & \multirow{-2}{*}{$1^{++}$} & $\gamma \jp$,$D\bar{D^*}$ & \multirow{-2}{*}{$B\rt KX(3872)$, $p\bar{p}$} & CDF, D0, \\

$X(3915)$& $3914\pm4$&$ 28^{+12}_{-14} $& $0/2^{++}$ &$\omega J/\psi$ & $\gamma\gamma\rt X(3915)$ &  \\

$Z(3930)$& $3929\pm5$&$ 29\pm10 $& $2^{++}$ & $D\bar{D}$ & $\gamma\gamma\rt Z(3940)$ &  \\

 &&&& $D\bar{D^*}$ (not $D\bar{D}$ &&\\
\multirow{-2}{*}{$X(3940)$} & \multirow{-2}{*}{$3942\pm9$} & \multirow{-2}{*}{$ 37\pm17 $} & \multirow{-2}{*}{$0^{?+}$} &  or $\omega J/\psi$)  & \multirow{-2}{*}{$\ee\rt \jp  X(3940)$} &   \\

$Y(3940)$& $3943\pm17$&$ 87\pm34 $&$?^{?+}$ & $\omega J/\psi$ (not
$D\bar{D^*}$) & $B\rt K Y(3940)$ &  BaBar \\
$Y(4008)$& $4008^{+82}_{-49}$&$ 226^{+97}_{-80}$ &$1^{--}$& $\pipi \jp$ &
$\ee$(ISR) & \\
$X(4160)$& $4156\pm29$&$ 139^{+113}_{-65}$ &$0^{?+}$& $D^*\bar{D^*}$
 (not $D\bar{D}$) & $\ee \rt \jp X(4160)$  & \\
$Y(4260)$& $4264\pm12$&$ 83\pm22$ &$1^{--}$&  $\pipi \jp$ & $\ee$(ISR)
&BaBar, CLEO    \\
$Y(4350)$& $4361\pm13$&$ 74\pm18$ &$1^{--}$&  $\pipi \psip$ & $\ee$(ISR)
& BaBar  \\
$X(4630)$& $4634^{+9}_{-11}$&$ 92^{+41}_{-32} $ &$1^{--}$&  $\Lambda_c^+\Lambda_c^-$ & $\ee $(ISR)
&      \\
$Y(4660)$& $4664\pm12$&$ 48\pm15 $ &$1^{--}$&  $\pipi \psip$ & $\ee $(ISR)
&      \\
$Z(4050)$& $4051^{+24}_{-23}$&$ 82^{+51}_{-29}$ & ? &
$\pi^{\pm}\chi_{c1}$ & $B\rt K Z^{\pm}(4050)$  &  \\
$Z(4250)$& $4248^{+185}_{-45}$&$ 177^{+320}_{-72}$ & ? &
$\pi^{\pm}\chi_{c1}$ & $B\rt K
Z^{\pm}(4250)$  &  \\
$Z(4430)$& $4433\pm5$&$ 45^{+35}_{-18}$ & ? & $\pi^{\pm}\psip$ & $B\rt K
Z^{\pm}(4430)$  &  \\
$Y_b(10890)$  & $10,890\pm 3$ & $55\pm 9$ &   $1^{--}$ &
$\pipi\Upsilon(1,2,3S)$
& $\ee\rt Y_b$ &  \\
\hline\hline
\end{tabular}%
\end{table*}

\section{Experimental environment}
The luminosity of the KEKB collider has been steadily increasing since the start of operation in 1999. Recently the highest luminosity ever reached in $e^+e^-$ collisions was achieved, exceeding $2.1\times 10^{34}$~cm$^{-2}$s$^{-1}$. This corresponds to more than 1.5 million $B\overline{B}$ pairs recorded by the Belle detector each day. The integrated luminosity of the total sample collected in 10 years of operation amounts to about 960~fb$^{-1}$. Most of the results presented here are obtained on a smaller data sample. 

The detector is configured within a 1.5 T superconducting
solenoid and iron structure surrounding the KEKB beams. The Silicon Vertex
Detector (SVD), situated just outside a cylindrical beryllium
beam-pipe, is used for precise reconstruction of decay vertices. Main charged particle tracking is provided by the Central Drift Chamber (CDC). The relative uncertainty of the measured transverse momentum is between 0.3\% and 1.1\% for tracks with momenta between 0.6 and 5~GeV/$c$, respectively. Particle identification is provided by $dE/dx$
ionization energy loss measurements in the CDC, Cherenkov light yield measurements in the Aerogel \v{C}erenkov counters (ACC) and time-of-flight counters (TOF). More than three standard deviations separation between kaons and pions is realized up to momenta of 3~GeV/$c$. Electromagnetic showers are detected with the Electromagnetic Calorimeter (ECL) that consists of an array of CsI($Tl$) crystals which also serve as the identification device of $e^{\pm}$ and photons. Muons and $K_L$ mesons are identified by arrays of resistive plate counters interspersed in the iron yoke. 

\section{\boldmath The $XYZ$ states with masses near 3940~MeV}
\begin{figure*}[t]
  \includegraphics[width=.99\textwidth]{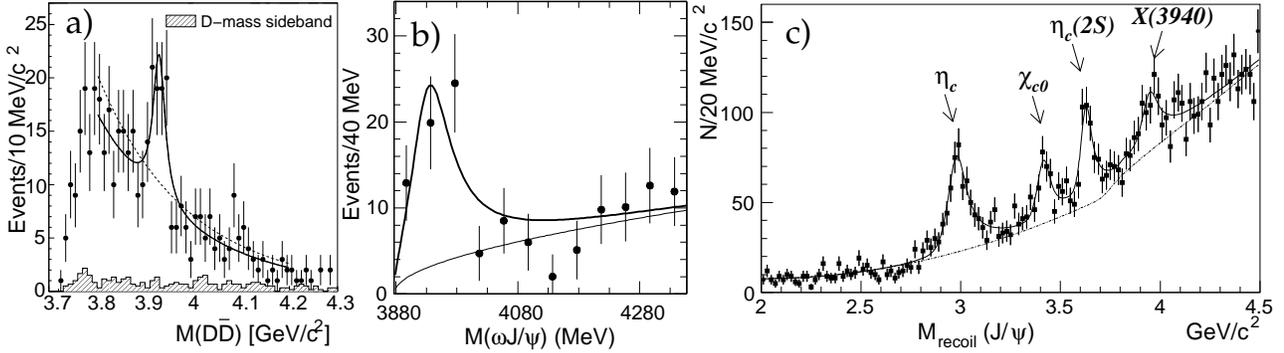}
  \caption{(a) Invariant mass of $D\overline{D}{}^{\ast}$ produced in two photon reactions \cite{Z3930_2005}. 
	   (b) Yield of $B$ mesons in $B\to K\omega J/\psi$ as a function of $M(\omega J/\psi)$ \cite{Y3940_2005}. 
	   (c) Spectrum of mass recoiling against the $J/\psi$ \cite{X3940_2005}. 
}
  \label{fig_2}
\end{figure*}

In 2005,  Belle reported observations of three states with masses near 3940~MeV (see Fig. \ref{fig_2}): the $X(3940)$, observed in the process $e^+e^- \to J/\psi X(3940)$, both in inclusive production and via the $X(3940)\to  D \overline{D}{}^{\ast}$ decay mode \cite{X3940_2005}; the $Y(3940)$, observed as a near-threshold enhancement in the $\omega J/\psi$ invariant mass distribution for exclusive $B\to K \omega J/\psi$ decays \cite{Y3940_2005}; and the $Z(3930)$, observed as a $D\overline{D}$ mass peak in two-photon collisions $\gamma\gamma\to D\overline{D}$ \cite{Z3930_2005}. The mass and width of the $Z(3930)$ are measured to be $(3929\pm 5\pm 2)$~MeV\footnote{Throughout this review, units are used in which $c=\hbar=1$.} and 
$29\pm 10 \pm 2$~MeV, respectively, and the product of the two-photon decay width and branching fraction of the $Z(3930)$ is found to be $\Gamma (Z(3930)){\cal B}(Z(3930)\to D\bar{D})=0.18\pm 0.05 \pm 0.03$~keV. 
An angular analysis showed that spin-2 assignment is strongly favored over spin-0 assignment of $Z(3930)$. All of the above mentioned properties of $Z(3930)$ match well to expectations for the $\chi'_{c2}$, a radial excitation of $2^3P_2$ charmonium state. 

\begin{figure}[t]
\includegraphics[width=0.49\textwidth]{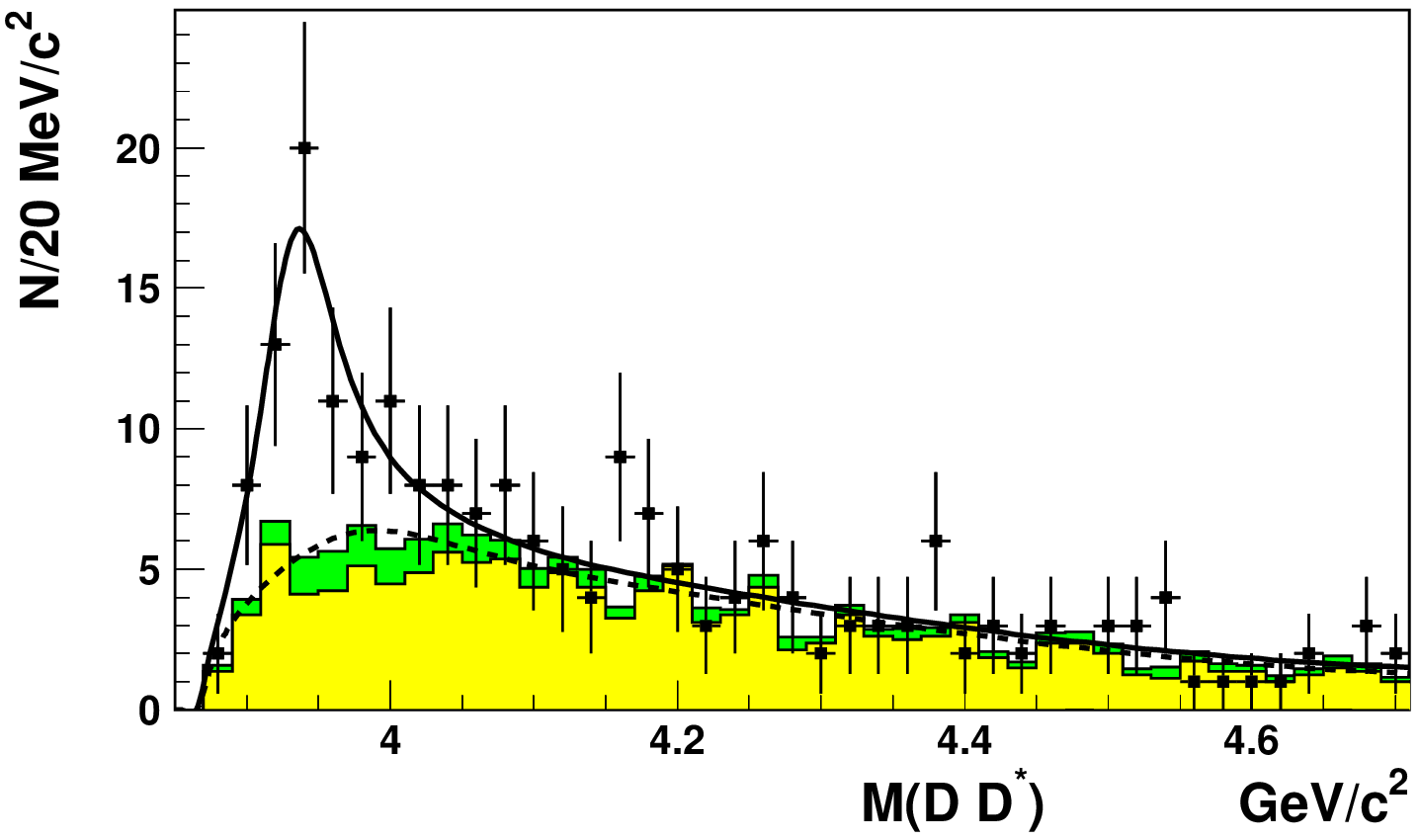}\\
\includegraphics[width=0.49\textwidth]{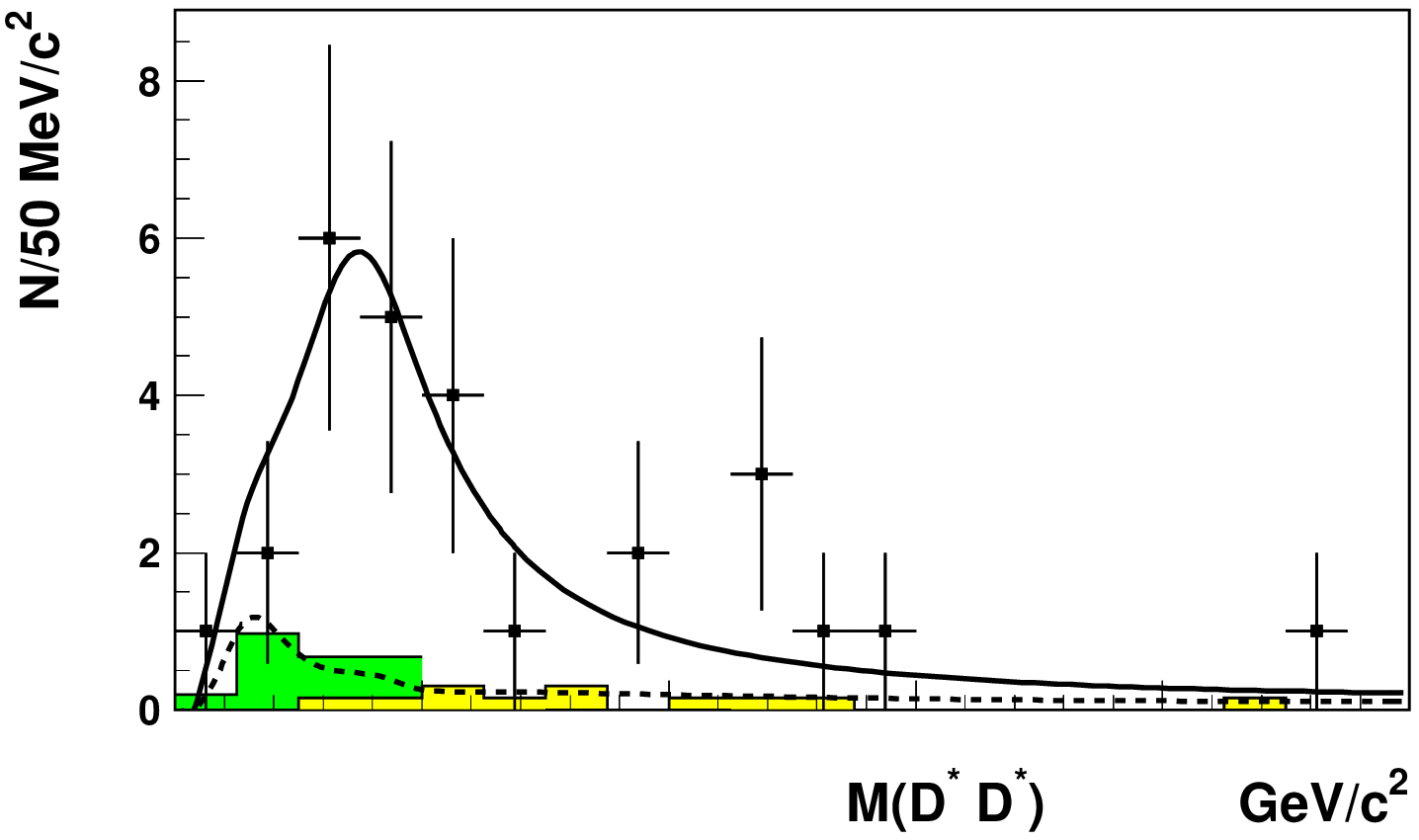}
\caption{(Top) The $M(D\overline{D}{}^{\ast})$ invariant mass distribution for the
  $e^+e^-\to J/\psi D\overline{D}{}^{\ast}$ process and (Bottom) the $M(D^{\ast}\overline{D}^{\ast})$ invariant mass distribution for the
  $e^+e^-\to J/\psi D^{\ast}\overline{D}^{\ast}$ process \cite{X4160}. Points with error bars correspond to the signal windows while histograms show the scaled sideband distributions.}
\label{fig5}
\end{figure}

Last year Belle confirmed the observation of $X(3940)\to D\overline{D}{}^{\ast}$ and reported an observation of a new charmoniumlike state, $X(4160)$, seen also in the double charmonium production process in $e^+e^-$ annihilation but decaying into $D^{\ast}\overline{D}^{\ast}$ \cite{X4160} (see Fig. \ref{fig5}). Neither are seen in the experimentally more accessible $D\overline{D}$ channel. Circumstantial evidence favour $J^{PC}=0^{-+}$ assignments for both states, leading to possible interpretation of these two states as $\eta_c(3S)$ and $\eta_c(4S)$ conventional charmoniums. However, the problem with this assignment is that potential models predict masses for these charmonium levels to be significantly higher than those measured for the $X(3940)$ and $X(4160)$ (given in Table \ref{tab_newstates}).

BaBar also observed a near-threshold $\omega J/\psi$ enhancement in the $B\to \omega J/\psi K$ decays \cite{BaBarY3940}, confirming the Belle result, but obtained lower mass and smaller width and reduced the uncertainty on each by a factor of around 3. The mass and width consistency between the $Y(3940)$ (seen in $\omega J/\psi$) and the $X(3940)$ (seen in $D\overline{D}{}^{\ast}$) suggests that the observed peaks are different decay modes of the same state. This explanation is unlikely since a 90\% CL lower limit of ${\cal B}(Y\to \omega J/\psi)/{\cal B}(Y\to D\overline{D}{}^{\ast})>0.71$ set by Belle searching for $Y(3940)\to D\overline{D}{}^{\ast}$ in $B\to  D\overline{D}{}^{\ast} K$ decays \cite{Zwahlen:2008su} is in a contradiction with a 90\% CL upper limit ${\cal B}(X\to \omega J/\psi)/{\cal B}(X\to D\overline{D}{}^{\ast})<0.58$ from a search for $X(3940)\to \omega J/\psi$ in $e^+e^-\to J/\psi \omega J/\psi$ annihilations performed by Belle \cite{X3940_2005}. 

The charmonium state with mass above open-charm mass thresholds is expected to dominantly decay to $D\overline{D}{}^{(\ast)}$, which for $Y(3940)$ were not observed yet. After taking into account also the fact the partial width to $\omega J/\psi$ for $Y(3940)$ is well above the measured partial widths for any of the observed hadronic transitions between charmonium states and lack of charmonium levels at the $Y(3940)$ mass one can conclude that the $Y(3940)$ is probably not a charmonium state. 

\subsection{\boldmath Observation of an enhancement in $\gamma\gamma\to \omega J/\psi$}
\begin{figure}[t]
\includegraphics[width=0.49\textwidth]{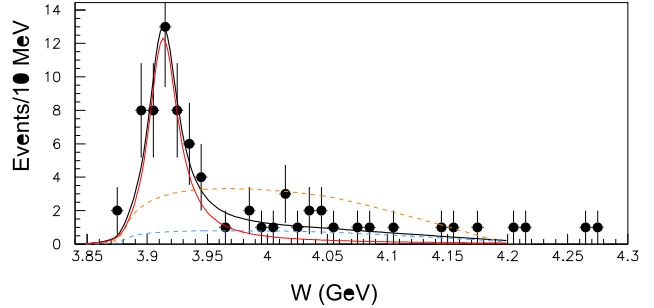}
\caption{The $\omega J/\psi$ mass distribution for selected $\gamma\gamma\to \omega J/\psi$ events. The solid curve shows the result of the fit that uses a phase-space-weighted, resolution-broadened S-wave Breit-Wigner function plus a smooth background function. The dashed curveshows the result of the fit with no Breit-Wigner term.}
\label{figX3915}
\end{figure}

Recently Belle reported an enhancement in the cross section for $\gamma\gamma\to \omega J/\psi$ in the $3.90$-$3.95$~GeV region \cite{X3915}, which may be related to one of the three states discussed above. 
The selected events with $\pi^+\pi^-\pi^0$ and $\ell^+\ell^-$ invariant masses within $\pm30$~MeV and $\pm25$~MeV wide windows around the nominal masses of the $\omega$ and $J/\psi$, respectively, are required to have a total transfer momentum balance of the final state particles less than 100~MeV in CM system. The two photon CM energy distribution of selected events is shown in Fig. \ref{figX3915}. An S-wave Breit-Wigner fit to this enhancement, gives preliminary results for the mass and width of this new state, denoted by $X(3915)$, of $M=3914\pm14\pm2$~MeV and $\Gamma = 28\pm12^{+2}_{-8}$~MeV, respectively, where the systematic errors are determined by varying the selection criteria and fitting procedure. The statistical significance of the signal is 7.1 standard deviations. The product of the two-photon decay width and the branching fraction for $X\to\omega J/\psi$ is determined to be $\Gamma_{\gamma\gamma}{\cal B}=69\pm16^{+7}_{-18}$~eV ($21\pm4^{+2}_{-5}$~eV) for assumed $J^P=0^+$ ($2^+$) assignment.

The preliminary value for the $X(3915)$ mass is about 2 standard deviations different from that of the $Z(3930)$ (see Table \ref{tab_newstates}). If one assumes that $X(3915)$ and $Z(3930)$ are different decay modes of the same state ($\chi_{c2}'$ charmonium), then a comparison of the products of the two-photon decay width and corresponding branching fractions yields the ratio of branching fractions ${\cal B}(\chi_{c2}'\to \omega J/\psi)/{\cal B}(\chi_{c2}'\to D\overline{D})\geq0.08$, which is quite large for charmonium. On the other hand, there is a good agreement between these preliminary results and the mass and width quoted by BaBar for the $Y(3940)$, which is also seen in $\omega J/\psi$. If $X(3915)$ is the same state as $Y(3940)$, then $J^{PC}=0^{++}$.

\section{\boldmath$X(3872)$}
In 2003 Belle discovered a charmoniumlike state $X(3872)$ as a narrow peak in the $\pi^+\pi^-J/\psi$ invariant mass distribution from $B^+\to K^+\pi^+\pi^-J/\psi$ decays \cite{belle:x3872}. The observation was later confirmed by CDF~\cite{cdf:x3872}, D0~\cite{d0:x3872} and BaBar~\cite{babar:x3872} experiments. 

Last year Belle presented new results on the $X(3872)$, produced in $B^+ \to X(3872) K^+$, $B^0 \to X(3872) K^0_S$ (first statistically significant observation) and $B^0 \to X(3872) K^+\pi^-$ decays where $X(3872) \to \pi^+ \pi^- J/\psi$~\cite{belle:X3872new}. The mass difference between the $X$ states produced in $B^{0,+}\to X(3872)K^{0,+}$ decays is found to be $\delta M=M_{XK^+}-M_{XK^0}=+0.18\pm0.89\pm0.26$~MeV, consistent with zero. The measurement of mass difference between the X states produced in neutral and charged $B$ decays is interesting since in the diquark-antidiquark model of the $X(3872)$, the $X$ produced in $B^+$ decays is interpreted as a $cu\overline{c}\overline{u}$ combination while the $X$ produced in $B^0$ decays is a $cd\overline{c}\overline{d}$ combination, which should differ in mass by $8\pm3$~MeV \cite{diquark3}. 
Combining the charged and neutral $B$ samples the mass of the $X(3872)$ is found to be $M_{X(3872)}^{\rm Belle}=3871.46\pm0.37\pm0.07$~MeV. The world average of all measurements that use the $\pi^+\pi^-J/\psi$ decay mode is $M_{X(3872)}^{\rm WA}=3871.46\pm0.19$~MeV (dominated by the updated measurement from the CDF \cite{cdf:x3872new}) which is just below of the $D^{\ast 0}\overline{D}{}^0$ mass threshold: $m_{D^{\ast 0}}+m_{D^0}=3871.81\pm0.36$~MeV \cite{PDG}. We also measured the ratio of branching ratios which is found to be $\frac{{\cal B}(B^0\to X(3872)K^0_S)}{{\cal B}(B^+\to X(3872)K^+)}=0.82\pm0.22\pm0.05$ and is consistent with unity. 
\begin{figure}[hb!]
\includegraphics[width=0.49\textwidth]{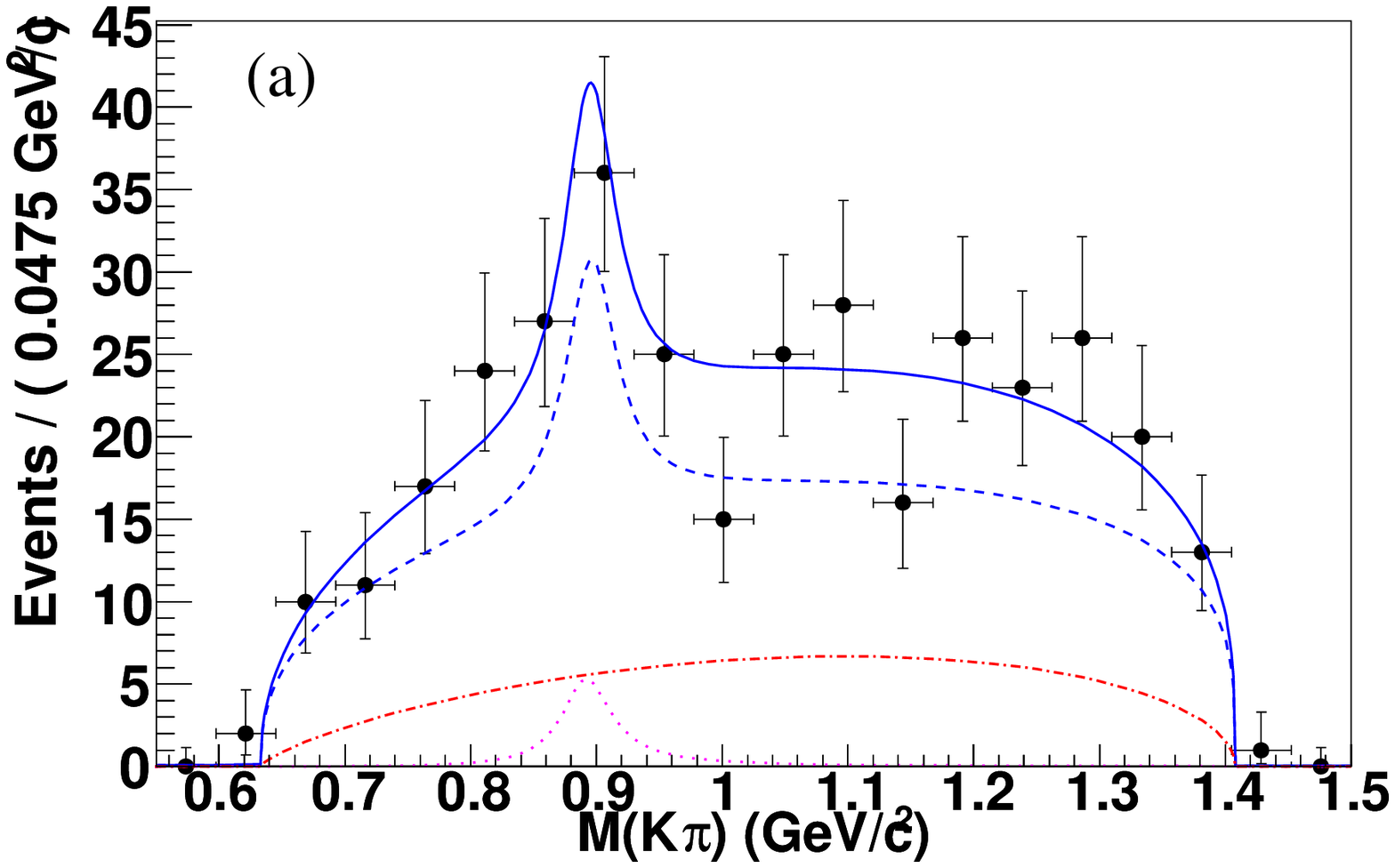}\\
\includegraphics[width=0.49\textwidth]{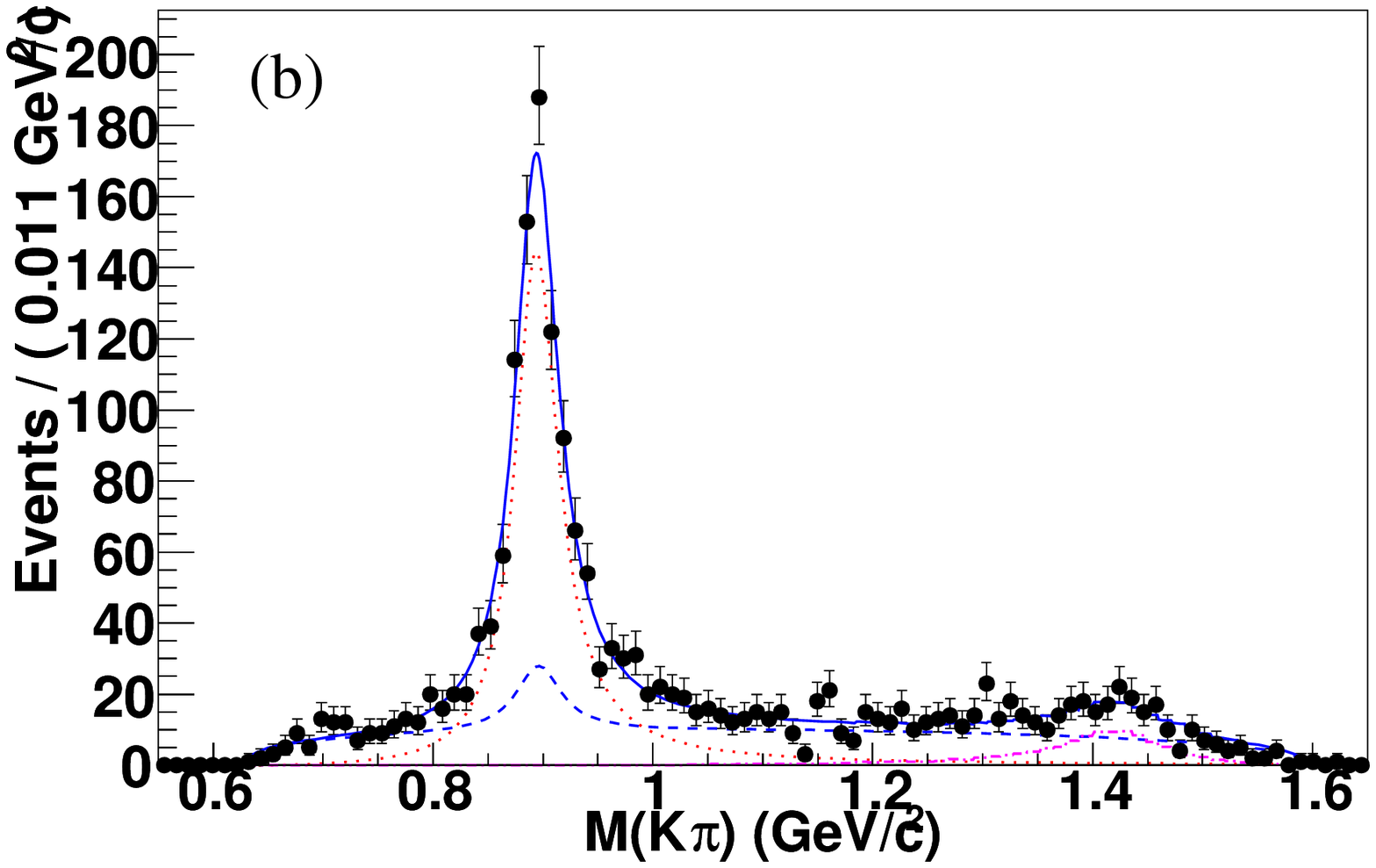}
\caption{The $K\pi$ mass spectrum for the $X(3872)$ region for $B^0\to\pi^+\pi^-J/\psi K^+\pi^-$ (a) and $B^0\to\psi(2S) K^+\pi^-$ decays. (a)
$B \to X(3872) (K^+ \pi^-)_{NR}$ is shown by the dash-dot red curve,
$B \to X(3872) K^{*}(892)^0$ by the dotted magenta curve,
and the background by the dashed blue curve.
(b) $B \to \psi(2S) K^{*}(892)^0$ is shown by the dotted red curve,
$B \to \psi(2S) K^{*}_2(1430)^0$ by the dash-dot magenta curve,
and the background by the dashed blue curve. 
}
\label{figX3872}
\end{figure}

We also performed a study of $X(3872)$ production in exclusive $B\to X(3872) K^+\pi^-$ decays \cite{belle:X3872new}. We observe a signal of $90\pm19$ events. A fit to the $M_{K\pi}$ distribution (see Fig. \ref{figX3872}) shows that non-resonant $K\pi$ production dominates over the $K^{\ast}(892)^{0}$ contribution. This is in contrast to what is observed in $B$ decays to conventional charmonium, such as $\psi(2S)$, in association with a $K\pi$, in which the resonant $K^{\ast}(892)^{0}$ contribution dominates, as seen in Fig. \ref{figX3872}. We measure ${\cal B}(B^0 \to X(3872) (K^+ \pi^-)_{NR})
\times {\cal B}(X(3872) \to J/\psi \pi^+ \pi^-) =
(8.1 \pm 2.0 ^{+1.1}_{-1.4} )\times 10^{-6}$
and we set the 90\% C.L. limit,
${\cal B}(B^0 \to X(3872)K^{*}(892)^0)
\times {\cal B}(X(3872) \to J/\psi \pi^+ \pi^-) < 3.4 \times 10^{-6}$.

The close proximity of the $X(3872)$ to the $D^{\ast 0}\overline{D}{}^0$ threshold motivated the interpretation of $X(3872)$ as a $DD^{\ast}$ molecule~\cite{molecular,molecular1,molecular2,molecular3,molecular4,molecular5,molecular6,molecular7,molecular8,molecular9} and the search for the $X(3872)\to D\overline{D}{}^{\ast}$ decays. 
In 2005 Belle showed a $6.4\sigma$ excess of events in the $\dnd
\pi^0$ invariant mass in the channel $B \to \dnd \pi^0 K$~\cite{belle:x3875} and BaBar reported an observation of $X(3875)$ decays to
\dndst~\cite{babar:x3875}. The masses of $X(3875)$ measured by both experiments are in agreement, however the weighted average significantly differs from the mass measured in the $\pi^+ \pi^- J/\psi$ decay mode. Last year Belle presented an updated study of near-threshold enhancement in the \dndst\ invariant mass spectrum in $B \to \dndst K$ decays~\cite{belle:x3875_3872}. The measured mass, $M=3872.6^{+0.5}_{-0.4}\pm0.4$~MeV, and width, $\Gamma=3.9^{+2.5}_{-1.3}{}^{+0.5}_{-0.3}$~MeV, are consistent with the current world average values for the $X(3872)$ in
the $\pi^+ \pi^- J/\psi$ mode. The obtained branching fraction and width are compatible with the values previously published
by Belle in Ref.~\cite{belle:x3875} for non-resonant $\dnd \pi^0$ decays. 


\section{Charged charmoniumlike states}
\subsection{\boldmath $Z^{\pm}(4430)$}
\begin{figure}[t]
\includegraphics[width=0.49\textwidth]{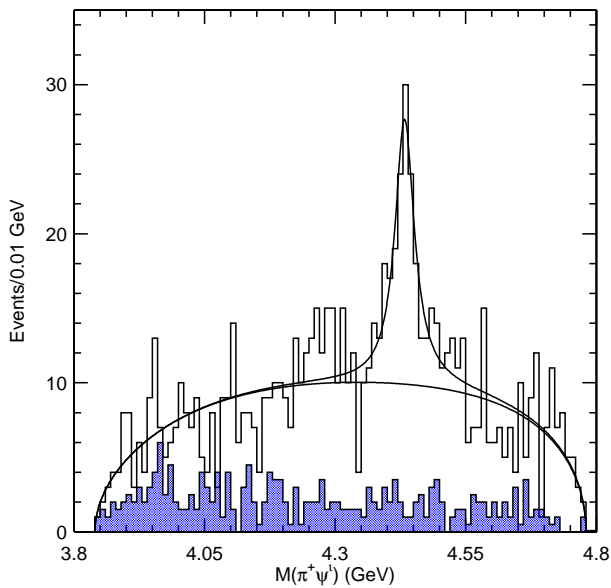}
\caption{
The $\pi^+\psi(2S)$ invariant mass distribution for $B\to K\pi^+\psi(2S)$ decays.
}
\label{figZ4430old}
\end{figure}

The first charged charmoniumlike state, denoted by $Z^+(4430)$, was discovered by Belle in 2007 in the $\pi^+\psi(2S)$ invariant mass distribution for $B\to\psi(2S)K\pi$ decays \cite{belle:z4430}. The $M_{\pi^+\psi(2S)}$ distribution of events after vetoing the $K^{\ast}(892)^0$ and $K^{\ast}_2(1430)^0$ peaks in $M_{K\pi}$ is shown in Fig.~\ref{figZ4430old}.  A fit to this distribution with an S-wave Breit-Wigner function for the signal and a phase-space-like function for the nonpeaking background gives $M=4422\pm4\pm1$~MeV and $\Gamma=44^{+17}_{-13}{}^{+30}_{-11}$~MeV. 

\begin{figure}[t]
\includegraphics[width=0.49\textwidth]{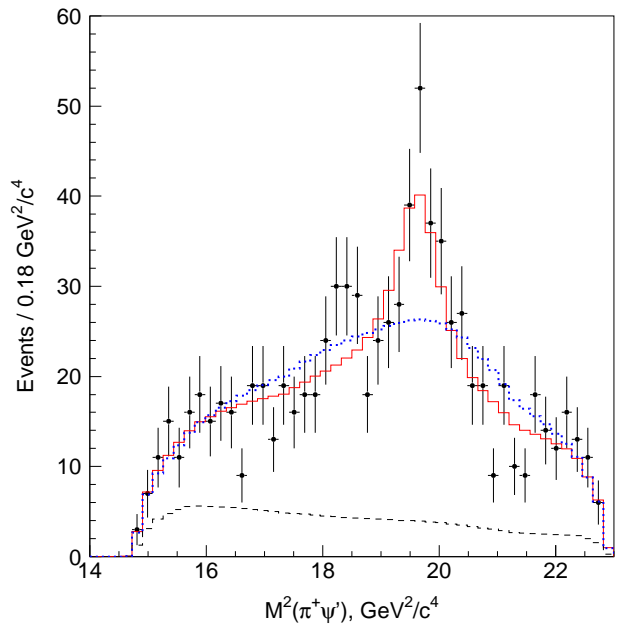}
\caption{
The $M^2_{\pi^+\psi(2S)}$ Dalitz plot projection with the prominent $K^{\ast}$ bands removed. The results of the fit with (without) $Z$ resonance are shown as a solid (dashed) histogram.
}
\label{figZ4430new}
\end{figure}

BaBar also performed a search for $Z^{\pm}(4430)\to J/\psi \pi^{\pm}$ and $Z^{\pm}(4430)\to \psi(2S) \pi^{\pm}$ in $B\to J/\psi(\psi(2S))K\pi^{\pm}$ decays~\cite{babar:z4430} and reported a 1.9 standard deviations signal with mass and width similar to Belle's. The significance of the signal increased to 3.1 standard deviations, after fixing the mass and width to the values obtained by Belle.
This year Belle reported a full Dalitz plot analysis of $B\to\psi(2S)K\pi$ decays~\cite{belle_z4430_dalitz} in order to check whether the dynamics in the $K\pi$ channel can cause mass structures in the $\pi\psi(2S)$ invariant mass distribution that have no relation to $\pi\psi(2S)$ dynamics. The fit to the $M^2_{K\pi}$ {\it vs.} $M^2_{\pi\psi(2S)}$ Dalitz plot distribution with the default fit model which includes all known low lying $K\pi$ resonances ($K^{\ast}(800)$, $K^{\ast}(892)$, $K^{\ast}(1410)$, $K^{\ast}_0(1430)$, $K_2^{\ast}(1430)$, $K^{\ast}(1680)$) fails to reproduce the narrow peak around 4.43~GeV in $M_{\pi\psi(2S)}$. The fit quality significantly improves after adding one $\pi\psi(2S)$ resonance ($Z(4430)$) to the default Dalitz model. Figure \ref{figZ4430new} shows the $M^2_{\pi\psi(2S)}$ Dalitz plot projection with superimposed results of the fits without and with $Z$ resonance. The fit with the $Z$ is favored over the fit with no $Z$ by 6.4 standard deviations. The fitted mass, $M=4443^{+15}_{-12}{}^{+19}_{-12}$~MeV, and width, $\Gamma=107^{+86}_{-43}{}^{+74}_{-56}$~MeV agree within systematic errors with the previous Belle result. A detailed systematic study is performed by considering a variety of other fit hypothesis. In all cases the significance of the $Z(4430)$ is found to be larger than 5.4 standard deviations. The product branching fraction from the Dalitz fit: ${\cal B^0}\to K^-Z^+\times {\cal B}(Z^+\to\pi^+\psi(2S))=(3.2^{+1.8}_{-0.9}{}^{+5.3}_{-1.6})\times 10^{-5}$ is not in strong contradiction with the BaBar 95\% CL upper limit of $3.1\times 10^{-5}$.

\subsection{\boldmath $Z^{\pm}(4050)$ and $Z^{\pm}(4250)$}
\begin{figure}[t]
\includegraphics[width=0.49\textwidth]{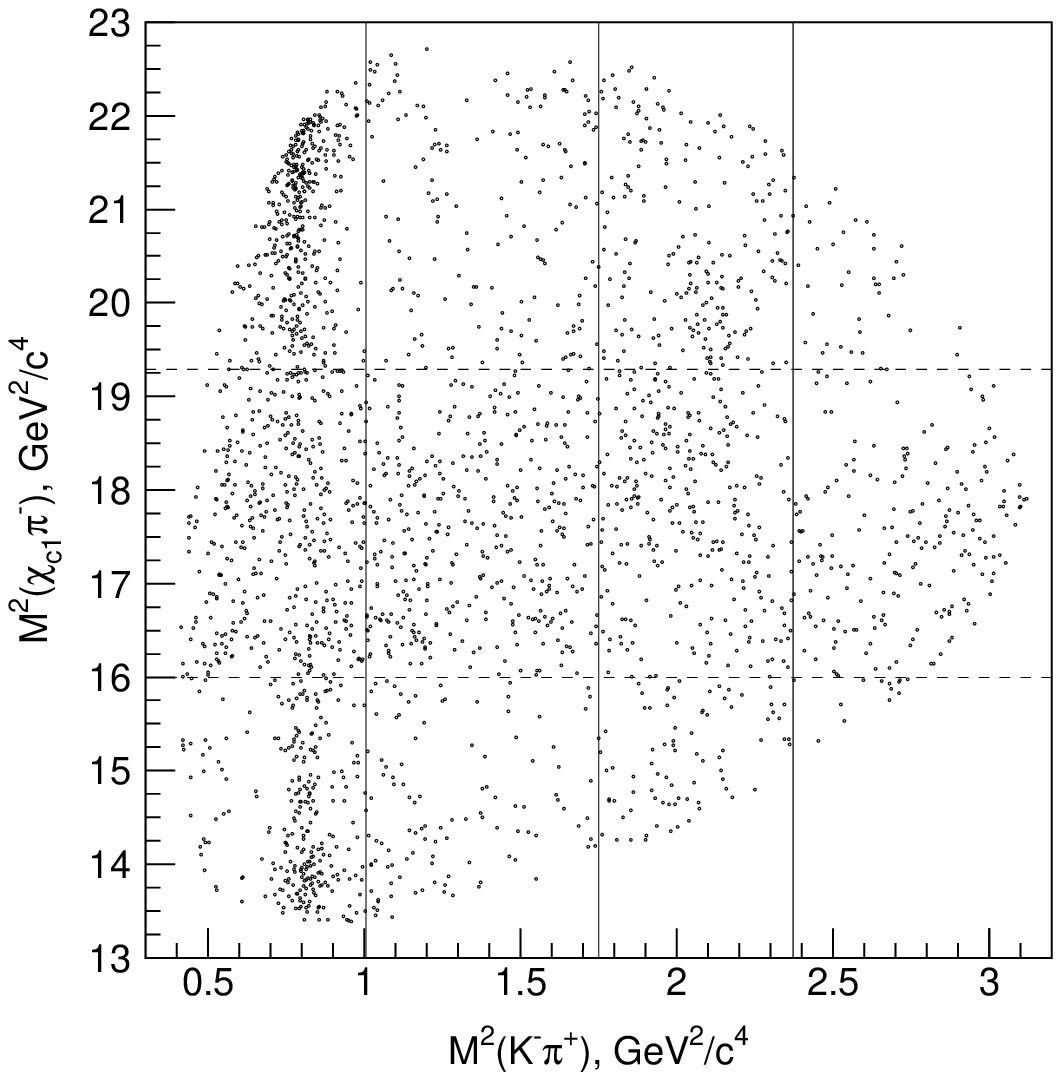}
\includegraphics[width=0.49\textwidth]{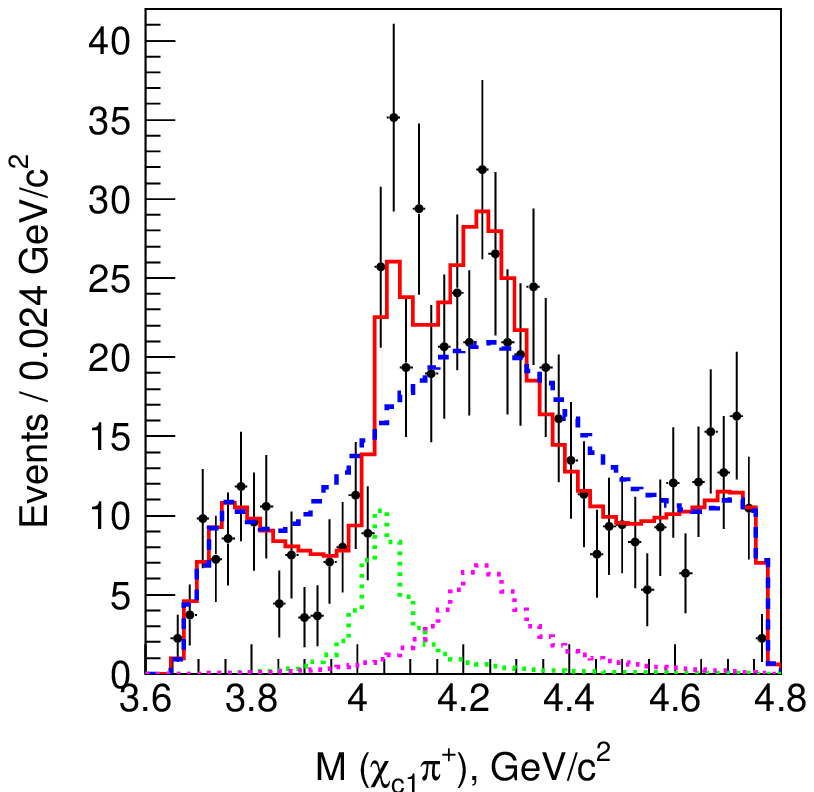}
\caption{
(Top) The $M^2_{K\pi}$ {\it vs.} $M^2_{\pi\chi_{c1}}$ Dalitz plot distribution for candidate $B\to K\pi^+\chi_{c1}$ events. (Bottom) The $M_{\pi\chi_{c1}}$ Dalitz plot projection for events from the $1.0~{\rm GeV}^2~<M^2_{K\pi}<1.75$~GeV$^2$ (second vertical band shown in top plot).
}
\label{fig2Z}
\end{figure}

In addition to the $Z^+(4430)$, Belle has discovered two new charged charmoniumlike states in the $\pi^+\chi_{c1}$ channel of $B\to K\pi^+\chi_{c1}$ decays~\cite{belle:z4050}. The Dalitz plot of selected events is shown in Fig. \ref{fig2Z} and exhibits some distinct features: two vertical bands at around 0.8~GeV$^2$ and 2~Gev$^2$, corresponding to two-body $K^{\ast 0}(892)\chi_{c1}$ and $K^{\ast}(1430)^0\chi_{c1}$ decays, respectively, and a distinct horizontal band around 17 Gev$^2$, indicating a structure in the $\pi^+\chi_{c1}$ channel. 
To describe the decay dynamics in the $K\pi$ channel we included the following eight $K^{\ast}$ resonances: $K^{\ast}(800)$, $K^{\ast}(892)$, $K^{\ast}(1410)$, $K^{\ast}_0(1430)$, $K_2^{\ast}(1430)$, $K^{\ast}(1680)$ and $K^{\ast}_3(1780)$. The result of the fit with $K\pi$ dynamics only is shown as a dashed histogram in Fig. \ref{fig2Z}. It fails to describe the observed $M^2_{\pi\chi_{c1}}$ distribution.
The fit with a single new $Z$ resonance in the ${\pi\chi_{c1}}$ channel is favored over the fit with only $K^{\ast}$ resonances and no $Z$ by more 10 standard deviations. Moreover, a fit with two $Z$ resonances is favored over the fit with only one $Z$ resonance by 5.7 standard deviations (the fit result with two $Z$ resonances included is shown in Fig.~\ref{fig2Z}). We also tried to fit with many different $K\pi$ resonance options, including the addition of new resonances with floating masses and widths, however no model with dynamics in the $K\pi$ channel only can describe the observed Dalitz plot. The fitted masses and widths of these two resonances are together with the product branching fractions given in Table \ref{tabZ2}. The product branching fractions are comparable to those of other charmoniumlike states as well as of the $Z^+(4430)$.
\begin{table}[t]
\caption{Masses, widths and product branching fractions of $Z^+(4050)$ and $Z^+(4250)$ observed in  $B\to K\pi^+\chi_{c1}$ decays.}
\label{tabZ2}
\begin{tabular}{l|cc}
& $Z(4050)^+$ & $Z(4250)^+$ \\\hline
$M$~[MeV] & $4051\pm14{}^{+29}_{-41}$ & $4248^{+44}_{-29}{}^{+180}_{-35}$\\
$\Gamma$~[MeV] & $82^{+21}_{-16}{}^{+47}_{-22}$ & $177^{+54}_{-39}{}^{+316}_{-61}$\\\hline
${\cal B}(\overline{B}^0\to Z^+K^-)\cdot$ & & \\
${\cal B}(Z^+\to \pi^+\chi_{c1})$ [$10^{-5}$] & $3.0^{+1.5}_{-0.8}{}^{+3.7}_{-1.6}$ & $4.0^{+2.3}_{-0.9}{}^{+19.7}_{-0.5}$
\end{tabular}
\end{table}

Nonzero electric charge makes these three states by definition exotic and excludes the possibility to interpret them as a charmonium or $c\overline{c}$-gluon hybrid mesons. They are prime candidates for a multiquark mesons. It is therefore important that the Belle results get confirmed by other experiments.

\section{\boldmath$1^{--}$ states via ISR}
ISR has proven to be powerful tool to search for $1^{--}$ states at $B$-factories, since it
allows to scan a broad energy range of $\sqrt{s}$ below the initial $e^+e^-$ CM energy, while 
the high luminosity compensates for the suppression due to the hard-photon emission.
\begin{figure*}[t]
\includegraphics[width=0.48\textwidth]{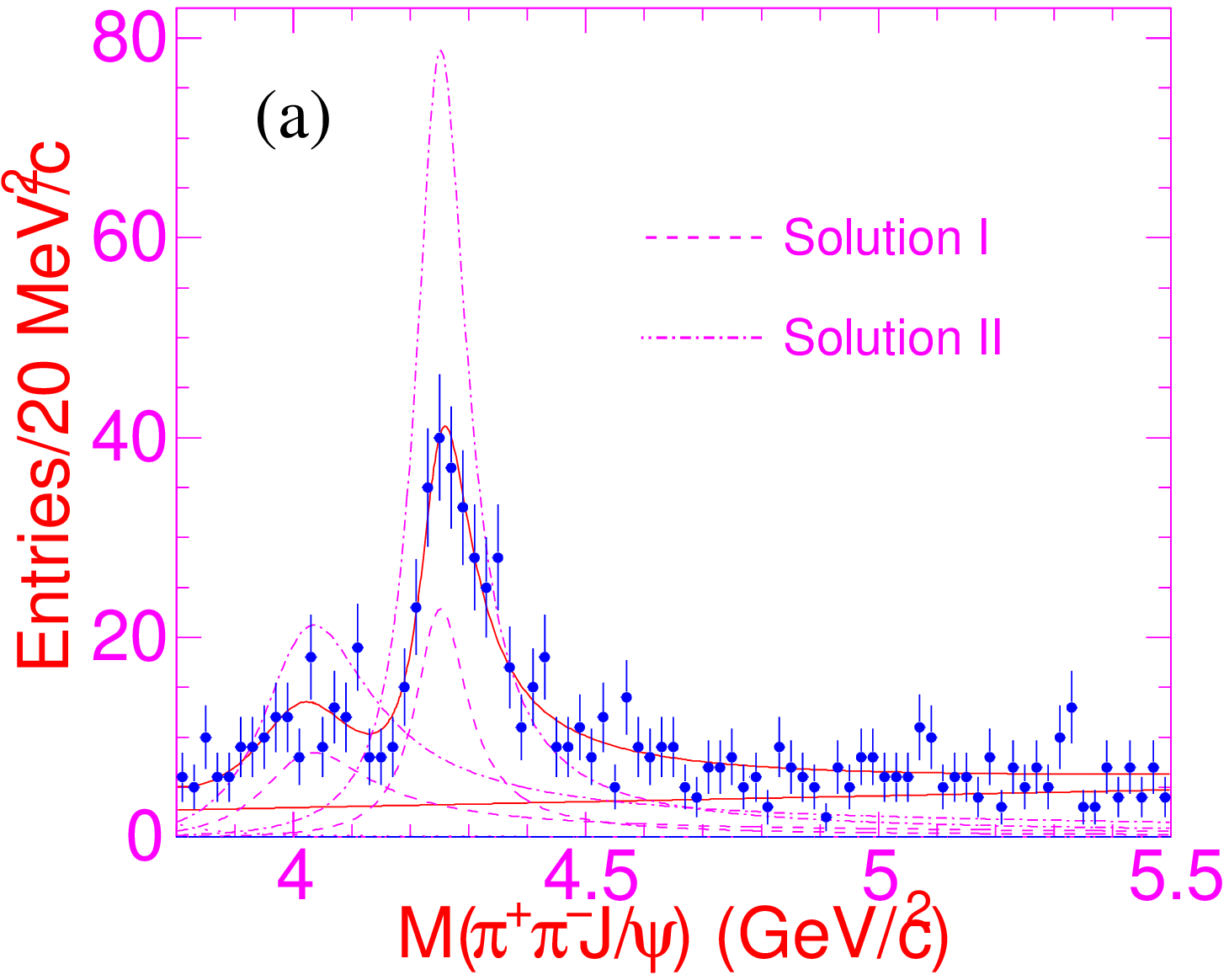}
\includegraphics[width=0.50\textwidth]{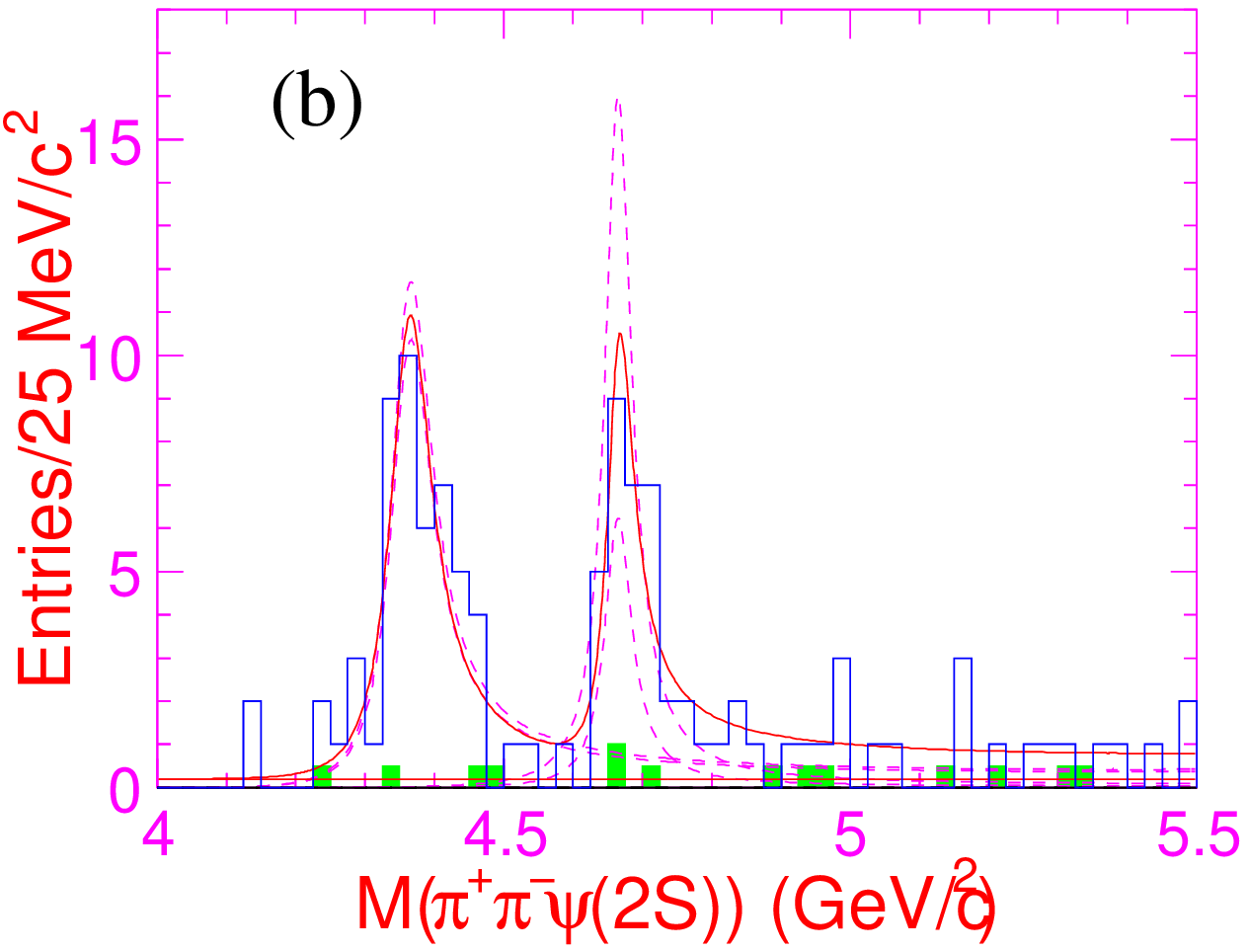}
\caption{The mass distributions for $\pi^+\pi^-J/\psi$ (a) and $\pi^+\pi^-\psi(2S)$ (b) combinations from $e^+e^-$ annihilation with the ISR photon. The curves in (a) and (b) show the fit results and contributions and contributions of individual resonances for constructive (Solution I) and destructive (Solution II) interference. 
}
\label{figY1MM}
\end{figure*}

\subsection{\boldmath $Y(4008)$ and $Y(4260)$}
First state, discovered with the ISR technique by BaBar, was $Y(4260)$ in the 
$e^+e^-\to \gamma_{\rm ISR}Y(4260)\to\gamma_{\rm ISR}\pi^+\pi^-J/\psi$ process \cite{babar:y4260}. 
Using the same method Belle confirmed the $Y(4260)$ state and in addition found another resonant 
structure, called $Y(4008)$ \cite{belle:y4260}. A fit using two interfering Breit-Wigner amplitudes 
to the $\pi^+\pi^-J/\psi$ invariant mass distribution describes the data better than a fit assuming one
resonance, especially for the lower-mass side of the $4.26$~GeV/$c^2$ enhancement (see Fig. 
\ref{figY1MM}(a)). Obtained masses and widths of the $Y(4008)$ and $Y(4260)$ are given in 
Table~\ref{tab_newstates}. 

\subsection{\boldmath $Y(4325)$ and $Y(4660)$}
The BaBar collaboration subsequently reported a similar structure in the
cross section for the ISR $e^+e^-$ annihilation process resulting in the $\pi^+\pi^-\psi(2S)$ 
final state~\cite{babar:y4360}. The mass and width of the state denoted as $Y(4325)$ 
are both significantly higher
than the values found for the $Y(4260)$. Belle performed a similar study on a larger data sample 
(673~fb$^{-1}$ for Belle compared to 272~fb$^{-1}$ for BaBar) and observed that the structure is formed
from two narrower peaks~\cite{belle:y4360}; one, in agreement with the BaBar study, is observed near 
4.36~$GeV^2$ and another, called $Y(4660)$, near 4.66~$GeV^2$. 
Fit to the $\pi^+\pi^-\psi(2S)$ distribution with a coherent sum of two Breit-Wigner amplitudes is shown
in Fig.~\ref{figY1MM}(b). Fitted masses and widths of the two states are given in Table~\ref{tab_newstates}. No sign was found either of $Y(4260)$ ($Y(4008)$) decay to $\pi^+\pi^-\psi(2S)$, or of $Y(4325)$
($Y(4660)$) decay to $\pi+\pi^-J/\psi$.

The nature of $Y$ states still reamins unclear. There is only one unassigned $1^{--}$ charmonium level
in this mass region, the $3^3D_1$ level, which might accommodate the $Y(4660)$, however there are no 
other available charmonium level in the spectrum for all other peaks discussed above. In addition,
these states, if considered to be conventional charmonium states, should decay mainly to 
$D^{(\ast)}\overline{D}{}^{(\ast)}$, however measurements of cross sections for exclusive open-charm 
final states in this energy range, perfromed by Belle~\cite{belle:dd,belle:dst,belle:4415}, 
BaBar~\cite{babar:dd} and 
CLEO-c~\cite{cleo:cs}, shown no evidence for peaking near the masses of the $Y$ states. 

\subsection{\boldmath $X(4630)$}
\begin{figure}[b]
\includegraphics[width=0.49\textwidth]{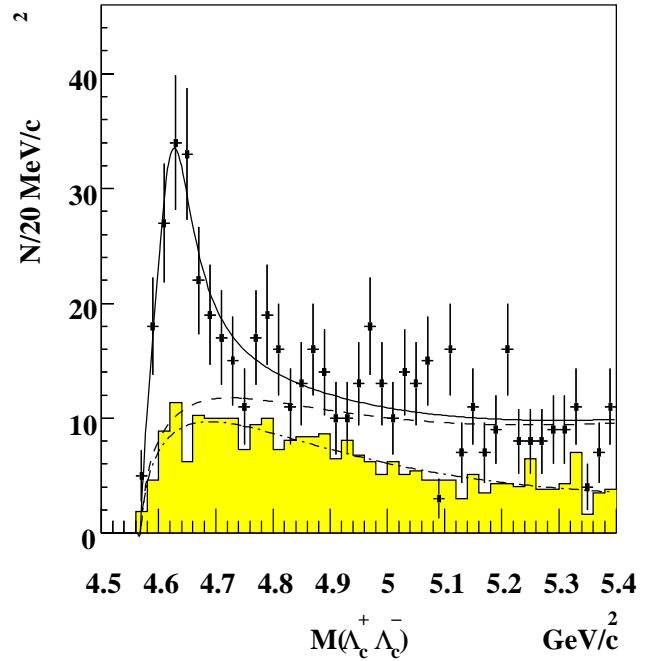}
\caption{The mass distributions for $\Lambda_c^+\Lambda_c^-$ combinations from $e^+e^-$ annihilation with the ISR photon. The curves show the result of the fit (solid line), threshold function (dashed line) and combinatorial background contribution (dashed-dotted line).
}
\label{figY2MM}
\end{figure}

The one exception is $e^+e^-\to \gamma_{ISR}\Lambda_c^+\Lambda_c^-$, for which Belle has reported a near-threshold
enhancement, called the $X(4630)$ (see Fig.~\ref{figY2MM}). The fitted mass and width are given in
Table~\ref{tab_newstates} and are consistent within errors with the mass and width of the $Y(4660)$ 
state. 

\section{\boldmath$XY$ counterparts in $b\overline{b}$ and $s\overline{s}$ systems}
An interesting question is whether or not there exist any $XY$ counterpart states in the $s\overline{s}$
and $b\overline{b}$ systems, predicted by many of the models proposed to explain the charmoniumlike 
$XYZ$ states. Some recent results, discussed below, indicate that this may be the case.

\subsection{\boldmath $Y(2175)$}
\begin{figure}[t]
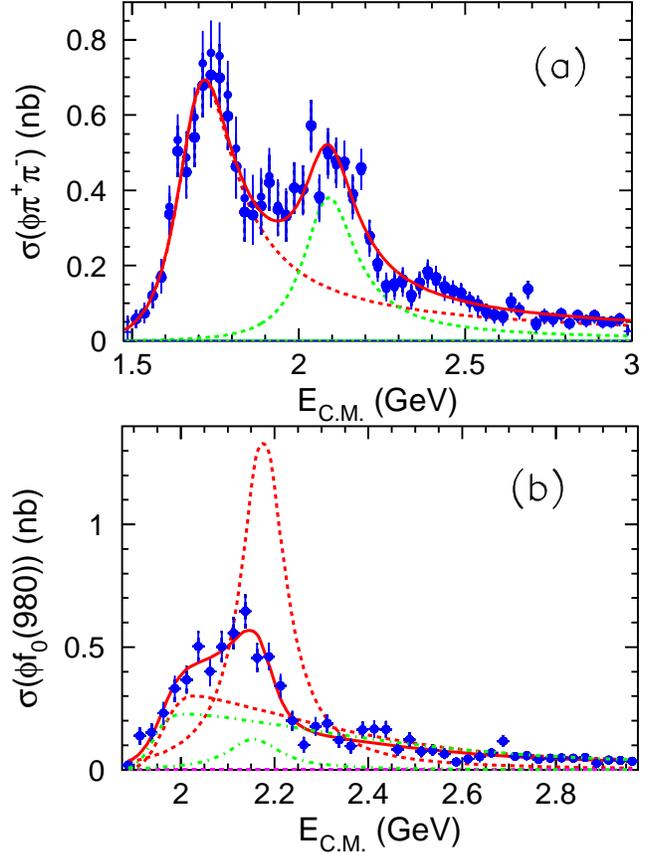

\includegraphics[width=0.49\textwidth]{fig10a.epsi}
\includegraphics[width=0.49\textwidth]{fig10b.epsi}
\caption{(a) The $e^+e^-\to\pi^+\pi^-\phi(1020)$ cross section distribution with superimposed fit result
with two coherent Breit-Wigner functions. (b) The $e^+e^-\to f_0(980)\phi(1020)$ cross section distribution with superimposed fit result with one resonance and a coherent non-resonant term.
}
\label{figYssbar}
\end{figure}

In the $s\overline{s}$ system one possible candidate is $Y(2175)$, a $1^{--}$ state, first observed by BaBar
in the ISR process $e^+e^-\to\gamma_{\rm ISR}f_0(980)\phi(1020)$~\cite{babar:y2175} and later also 
confirmed by BES~\cite{bes:y2175}. At Belle, we also measured the $\pi^+\pi^-\phi(1020)$ and $f_0(980)\phi(1020)$ cross sections for the ISR processes $e^+e^-\to\gamma_{ISR}\pi^+\pi^-\phi(1020)$ and 
$e^+e^-\to\gamma_{ISR}f_0(980)\phi(1020)$ with CM energy ranging from 1.5 to 3.5~GeV~\cite{belle:y2175}.
Figure~\ref{figYssbar} shows observed $\pi^+\pi^-\phi(1020)$ and $f_0(980)\phi(1020)$ cross section 
distributions. The former is fitted with two coherent Breit-Wigner functions and the latter with one
Breit-Wigner function interfering with a non-resonant continuum function. The mass and width of the high
mass peak, corresponding to $Y(2175)$, are found to be $M=2079\pm13^{+79}_{-28}$~MeV and $\Gamma=192\pm23^{+25}_{-61}$~MeV, which are consistent with the previous measurements. First measurements of mass and width
are reported for the low mass peak in the $\pi^+\pi^-\phi(1020)$ cross section distribution, which 
corresponds to the $\phi(1680)$. They are found to be $M=1689\pm7\pm10$~MeV and $\Gamma=211\pm14\pm19$~MeV. The widths of the $\phi(1680)$ and $Y(2175)$ are found to be quite similar and both are at the 200~MeV level. This may suggest that the $Y(2175)$ is an excited $1^{--}$ $s\overline{s}$ state. Since the
$f_0(980)$ is thought to have a large $s\overline{s}$ component, $Y(2175)\to f_0(980)\phi(1020)$ can be
viewed as an open-flavor decay as opposed to $Y(4260)\to \pi^+\pi^-J/\psi$, which is a hadronic 
transition. Studies of $Y(2175)$ in other decay modes are needed to distinguish, whether $Y(2175)$ is a
conventional $s\overline{s}$ state or an $s$-quark counterpart of the $Y(4260)$.

\subsection{\boldmath $Y_b(10890)$}
\begin{figure}[t]
\includegraphics[width=0.495\textwidth]{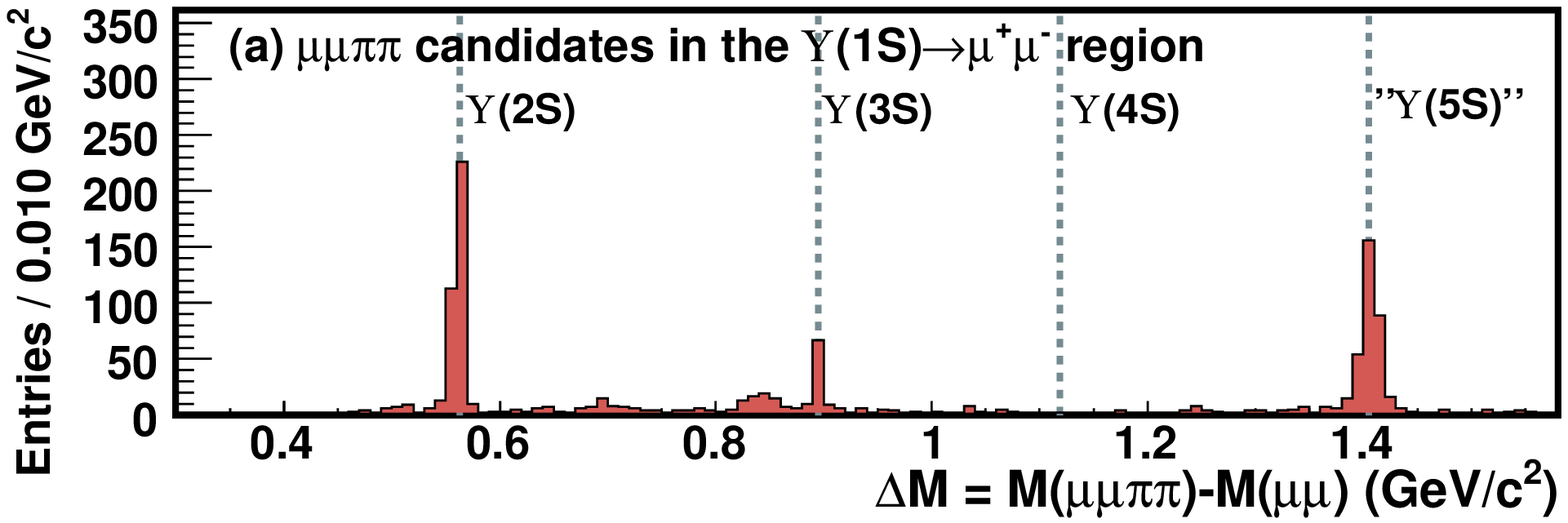}
\includegraphics[width=0.495\textwidth]{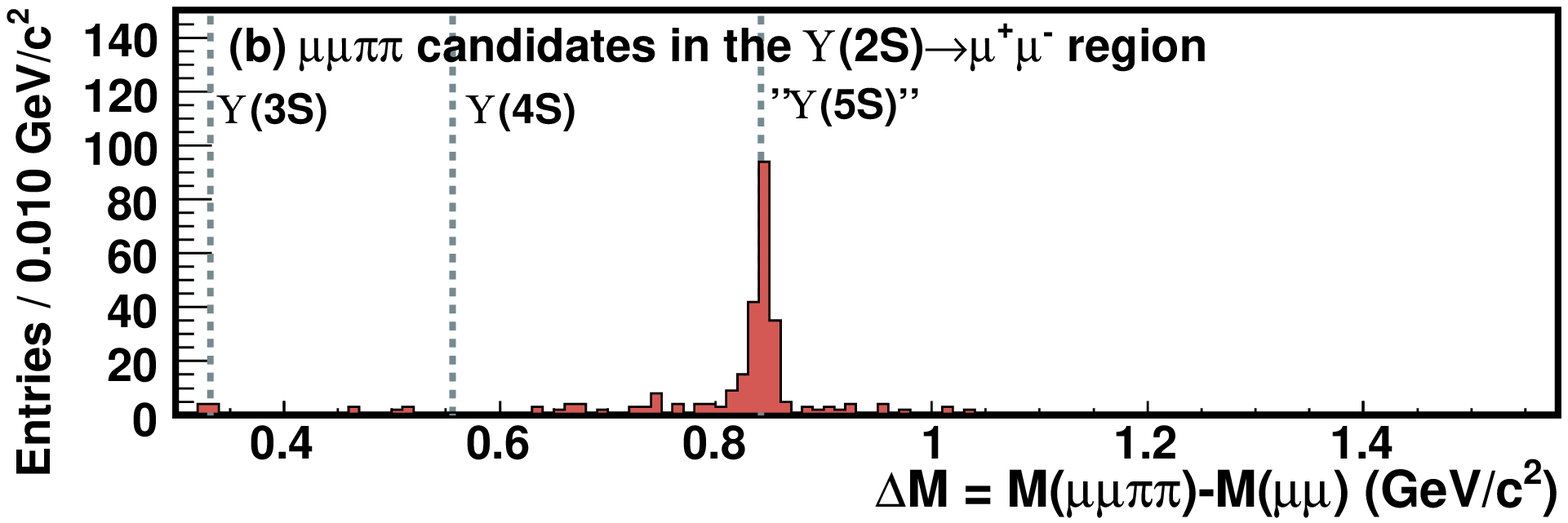}
\caption{The $\Delta M = M(\mu^+\mu^-\pi^+\pi^-)-M(\mu^+\mu^-)$ mass difference distributions for events
with $M(\mu^+\mu^-)=\Upsilon(1S)$ (a) and $M(\mu^+\mu^-)=\Upsilon(2S)$ (b). Vertical dashed lines indicate
the expected $\Delta M$ values for the $\Upsilon(nS)\to \pi^+\pi^-\Upsilon(1,2S)$ transitions.
}
\label{fig5strans}
\end{figure}

Belle and BaBar measured the partial decay widths of the order of few keV for $\Upsilon(4S)\to\pi^+\pi^-\Upsilon(1S)$ as 
well as $\pi^+\pi^-\Upsilon(2S)$~\cite{belle:y4strans,babar:y4strans} and found that are similar to 
those from dipion transitions from the $\Upsilon(3S)$ to the $\Upsilon(2S)$ and 
$\Upsilon(1S)$~\cite{PDG}. The same measurement was also perfromed by Belle on a data sample at the CM
energy of 10.87~GeV, corresponding to the $\Upsilon(5S)$, and found huge signals for 
$\pi^+\pi^-\Upsilon(1S)$, $\pi^+\pi^-\Upsilon(2S)$ and $\pi^+\pi^-\Upsilon(3S)$ (see Fig.~\ref{fig5strans}). Assuming the observed signal events are due solely to the $\Upsilon(5S)$ resonance, than the 
corresponding partial widths are found to be in the range (0.52--0.85)~MeV, more than two orders
 of magnitude larger than the corresponding partial widths for $\Upsilon(4S)$, $\Upsilon(3S)$ and 
$\Upsilon(2S)$ decays to $\pi^+\pi^-\Upsilon(1S)$. A possible explanation is a $b\overline{b}$ 
counterpart to the $Y(4260)$, denoted as $Y_b$~\cite{Hou:Yb}, which may overlap with the 
$\Upsilon(5S)$. Alternative explanations include a nonperturbative approach for the calculation of the 
decay widths of dipion transitions of heavy quarkonia~\cite{Simonov:Yb}, the presence of final state 
interactions~\cite{Meng:Yb}, or the existence of a tetraquark intermediate state~\cite{Karliner:Yb}. 

\begin{figure}[t]
\includegraphics[width=0.495\textwidth]{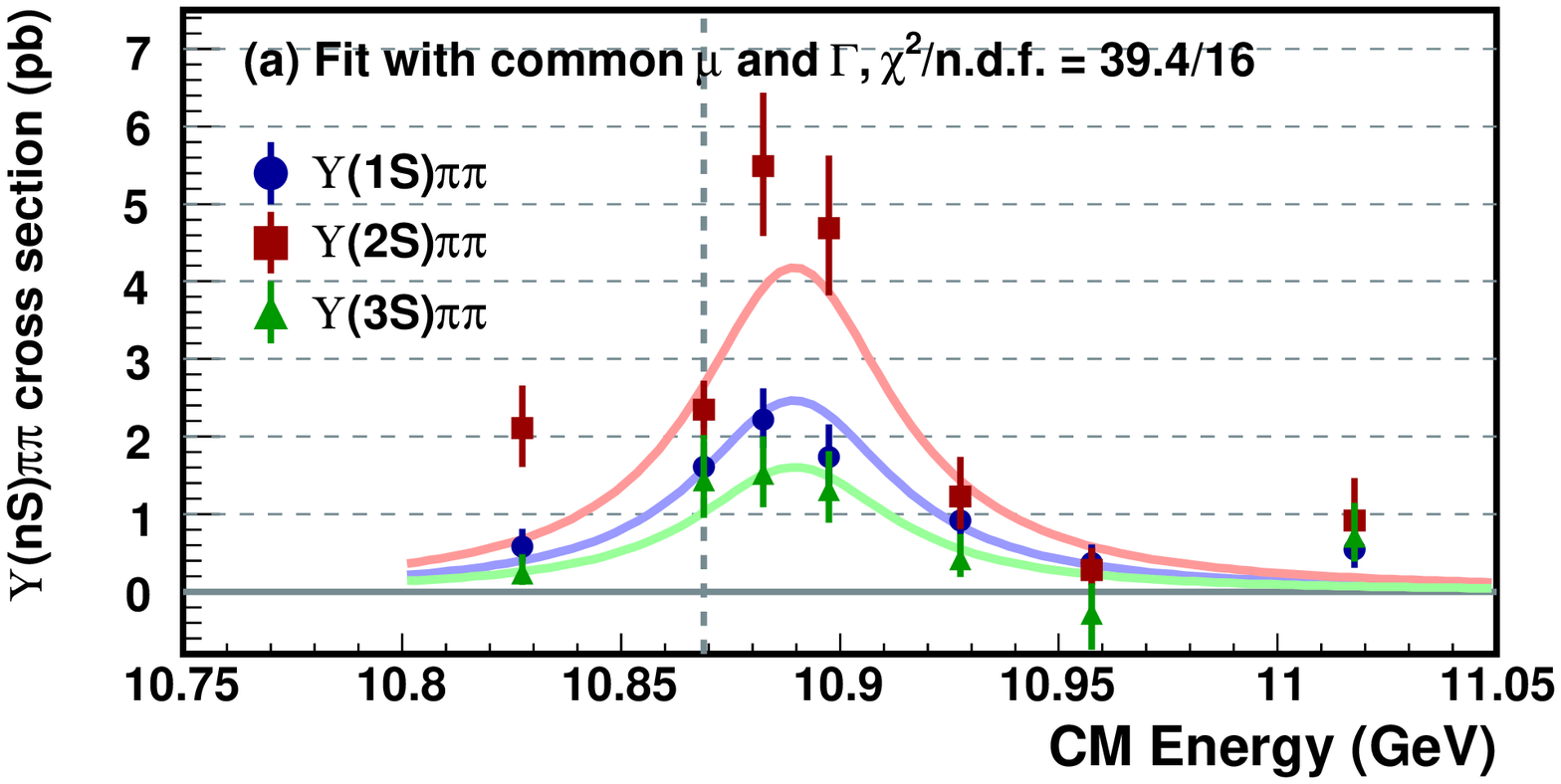}
\includegraphics[width=0.495\textwidth]{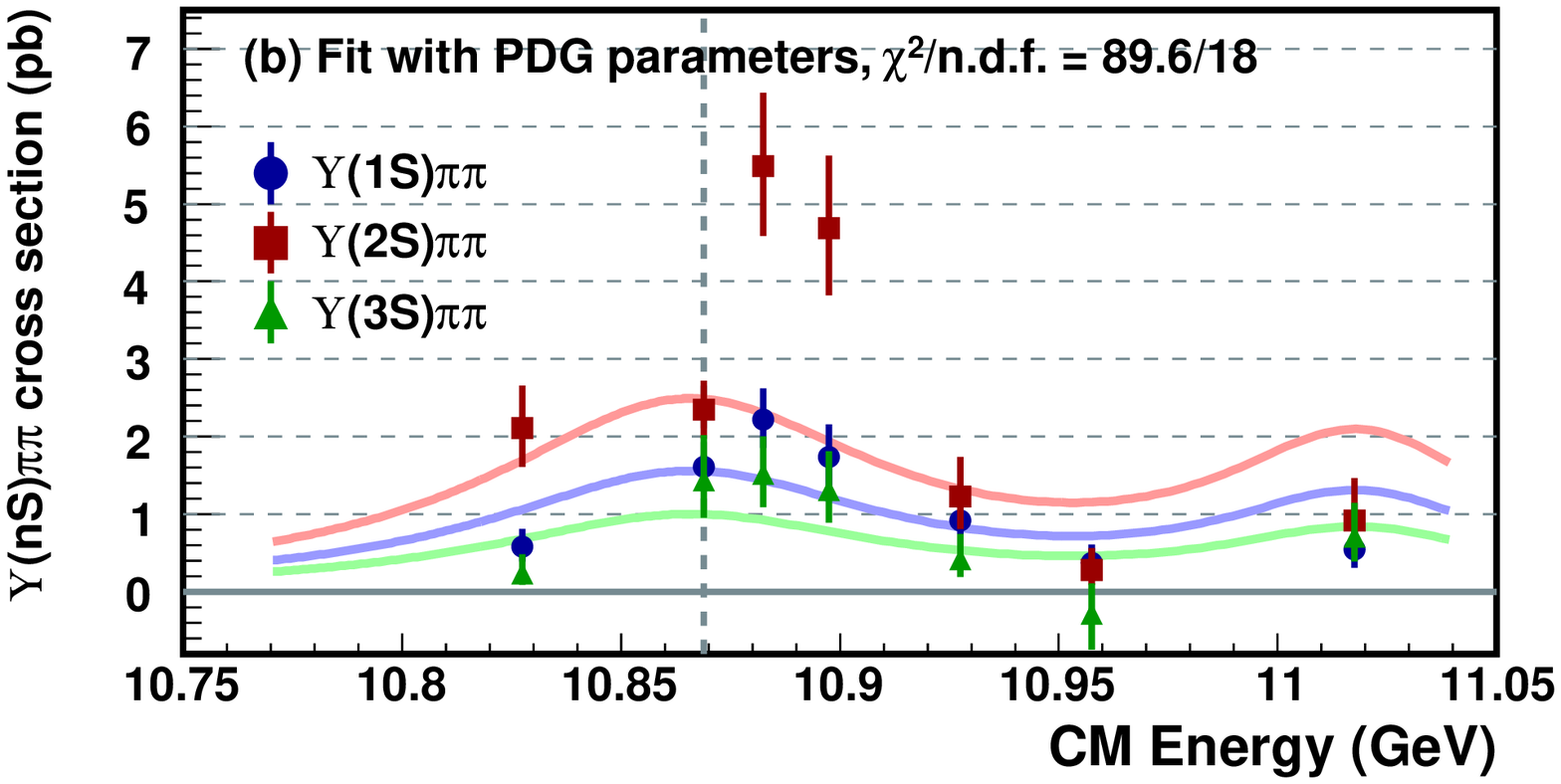}
\caption{The CM energy dependent cross section for $e^+e^-\to\pi^+\pi^-\Upsilon(nS)$ ($n=1,2,3$) processes. The results of the fits
which are shown as smoothed curves with (a) free common mean and width and (b) with fixed mean and width to the PDG $\Upsilon(5S)$ and
$\Upsilon(6S)$ parameters.
}
\label{figyb}
\end{figure}

In order to distinguish these hypotheses Belle performed a measurement of the energy dependence of 
the cross sections for $e^+e^-\to \pi^+\pi^-\Upsilon(nS)$ ($n=1,2,3$) at energies around 10.87~GeV~\cite{belle:Yb}. 
Peaks are found in all three channels at around 10.899~GeV. The preliminary values for the peak mass and width, obtained by performing a fit with a common 
Breit-Wigner function to the measured $\pi^+\pi^-\Upsilon(nS)$ cross section distribution 
(shown in Fig.~\ref{figyb}), are found to be $M=10889.6\pm1.8\pm1.6$~MeV and 
$\Gamma=54.7^{+8.5}_{-7.2}\pm2.5$~MeV. A fit using the PDG resonance parameters for the $\Upsilon(5S)$ and $\Upsilon(6S)$~\cite{PDG}, 
shown in Fig.~\ref{figyb}(b), fails to describe the observed $\pi^+\pi^-\Upsilon(nS)$ cross section. 

\section{Summary}
The Belle experiment at the KEKB collider provides an excellent environment for charmonium spectroscopy. As a result,
many new particles, summarized in Table~\ref{tab_newstates}, have been discovered during the 
ten year operation of the Belle detector. 

We have reported on a new Belle $X(3915)\to\omega\jpsi$ mass peak in $\gamma\gamma\to\omega\jpsi$. The measured
$X(3915)$ mass and width are in agreement with the BaBar's mass and width values for the $Y(3940)\to\omega\jpsi$
resonance seen in $B\to K\omega\jpsi$ decays. The problem with the interpretation of this state as a conventional
charmonium is its large partial decay width to $\omega\jpsi$.

The mass of the $X(3872)$ produced in $B^0\to K^0X(3872)$ and $B^+\to K^+X(3872)$ are found to be consistent within 
around 0.9~MeV. The mass of the $X(3872)$ decaying into $DD^{\ast}$ final state is consistent with that from 
$\pi^+\pi^-\jpsi$ within uncertainties. Another feature of $X(3872)$ is that the non-resonant $K^+\pi^-$ contribution
in $B^0\to K^+\pi^- X(3872)$ decay dominates over the resonant $K^{\ast 0}$ contribution, which is not the case for
conventional charmonium states.

There are too many $1^{--}$ $Y$ states and too few unassigned charmonium levels between 4.0~GeV and 4.7~GeV. In addition
measurements of cross sections for exclusive open-charm final states in this energy range show no evidence for peaking
near the masses of the $Y$ states, except for $e^+e^-\to \Lambda_c^+\Lambda_c^-$, which has a threshold peak near the
$Y(4660)$ peak mass.

Three charged $Z$ states were reported by Belle. Their quark content is $c\overline{c}u\overline{d}$ which makes them
manifestly exotic. It is important that the Belle results get confirmed by other experiments.

Belle reported evidence for possible analogue of $s\overline{s}$ and $b\overline{b}$ systems. The $Y_b(10890)$
found in $\pi^+\pi^-\Upsilon(nS)$ decays has different structure than $\Upsilon(5S)$.


\bigskip 

\end{document}